\documentclass[aps,prb,twocolumn,floatfix,citeautoscript]{revtex4-1}
\usepackage{amsmath}
\usepackage{amssymb}
\usepackage{graphicx}
\usepackage[caption=false]{subfig}
\usepackage{bm}

\renewcommand{\hat}[1]{#1}

\begin{document}

\title{Localization and self-trapping in driven-dissipative polariton condensates}
\author{F. Bello and P. R. Eastham}
\affiliation{School of Physics and CRANN, Trinity College Dublin, Dublin 2, Ireland}

\begin{abstract}
We study driven-dissipative Bose-Einstein condensates in a two-mode
Josephson system, such as a double-well potential, with asymmetrical
pumping. We investigate nonlinear effects on the condensate
populations and mode transitions. The generalized Gross-Pitaevskii
equations are modified in order to treat pumping of only a single
mode. We characterize the steady-state solutions in such a system as
well as criteria for potential trapping of a condensate mode. There
are many possible steady-states, with different density and/or phase
profiles. Transitions between different condensate modes can be
induced by varying the parameters of the junction or the initial
conditions, or by applying external fields.
\end{abstract}

\maketitle
\section{Introduction}\label{sec1}
Structures which enable strong coupling between excitons and photons
support hybrid light-matter quasiparticles known as
exciton-polaritons. As with cold atoms, above a certain density a
population of polaritons may condense into a single mode, and so form
a Bose-Einstein condensate (BEC)\cite{byrnes14,kavokin07}. Such
macroscopic quantum coherence leads to a number of nonlinear dynamical
effects, apparent in the condensate density, in trapped atomic BECs
and superconducting Josephson junctions. These include Landau-Zener
tunneling, population inversion, and self-trapping
\cite{smerzi97,raghavan99,han96,yu02,liu03,carlo10,ye08}. The realization of
analogous effects for exciton-polariton condensates is important for
optical communications technologies demanding ultrafast optical
switches, long-range coherence, as well as picosecond quantum
computing processes\cite{kasprzak06,winkler13}. In this direction,
advances in fabrication have led to successful demonstrations of
Josephson oscillations as well as self-trapping in micropillar
cavities\cite{lagoudakis10,abbarchi13}. The interplay between
interactions and coherence, central to such phenomena, also has an
important role in the formation of localized structures such as gap
solitons\cite{cerdamendez13,ostrovskaya13} and
vortices\cite{keeling08,ostrovskaya12}.

The polariton system differs from the atomic one because polaritons
decay, with a lifetime which is typically on the order of a few
picoseconds. The polariton condensate is thus a non-equilibrium
steady-state established by a balance of loss and gain, with the
latter provided by stimulated scattering from an incoherent
reservoir. This creates an interesting phenomenology, blending that
of equilibrium condensates, which are dominated by interactions, with
that of systems such as lasers, which are dominated by dissipation.
For example, in two-mode systems the steady-state can have density
oscillations, which correspond to the a.c. Josephson
effect\cite{wouters08,eastham08,lagoudakis10,read10,borgh10,pavlovic13,eastham14}. Such
oscillations also occur in the dissipation-dominated regime, where
they reflect the coexistence of two condensates of different frequencies, as in a multimode
laser\cite{eastham08}. Indeed at the qualitative level many important
phenomena, like gap solitons\cite{cerdamendez13,ostrovskaya13},
vortices\cite{keeling08}, and condensate
localization\cite{wouters08-2,ostrovskaya12,roumpos10}, occur in both
interaction-dominated and dissipation-dominated condensates.

In this paper we reexamine how the combination of gain, loss, and
interactions affects the behavior of a polariton condensate in the
simple two-state model of a Josephson
junction\cite{wouters08,eastham08,borgh10}. The two states could
correspond to neighboring local minima of a double-well potential in a
coupled photonic molecule\cite{abbarchi13}, or in a planar structure
at a suitable point in a disorder potential\cite{lagoudakis10}. They
could, alternatively, correspond to two polarization states of a
single trap mode\cite{shelykh08}. Previous work on the incoherently
pumped system has assumed that both modes are
pumped\cite{wouters08,eastham08,read10,borgh10,eastham14}. Here we
suppose that the single-particle modes are approximately symmetrical,
but only one is pumped. We find that this model has a rich phase
diagram, with many different steady-states.  These include states
where the density is symmetrical in the two wells, as well as states
where the condensate becomes trapped in the pumped well. The
appearance of asymmetrical states is similar to the phenomenon of
macroscopic quantum self-trapping, as previously analyzed for
polaritons without gain and loss\cite{shelykh08}, and for resonant
pumping\cite{sarchi08,cao16}. In these cases, however, the asymmetry
is caused by strong interactions, which prevent a complete transfer of
population between the wells. We show that this requirement can be
relaxed in the non-equilibrium case, and that the nonlinear gain, as
well as interactions, can induce a form of self-trapping.

Condensation in photonic molecules formed from two micropillar
resonators with asymmetrical pumping has been investigated by Galbiati
et al.\ \cite{galbiati12}. However, although in this experiment the
photonic molecule was symmetrical, the pumping led to a large detuning
between the states of each resonator. This was caused by the
interaction with the high energy excitons created by the pump. In this
large detuning regime the Josephson coupling has a negligible effect,
and our calculations do not apply. Our predictions could, instead, be
tested by using asymmetrical photonic molecules, constructed such that
the single-particle states come into resonance when they are
blueshifted by the interaction with the exciton reservoir.

\section{Model}\label{sec2}

We model the condensate at the mean-field level, using the complex
Ginzburg-Landau equation for the macroscopic
wavefunction\cite{keeling08}. This generalizes the Gross-Pitaevskii
equation (or nonlinear Schr\"odinger equation) to include the loss of
polaritons due to their decay into photons, and the gain due to
scattering into the condensate from exciton states at high
energies. In the case of a polariton condensate these processes can be
treated using local terms, because the exciton mass is very large and
gain diffusion is negligible.

We consider the complex Ginzburg-Landau theory for a double-well
potential in the two-mode limit, where the polaritons can occupy the
ground states of a well on the left or the right of a barrier. The
amplitudes of the macroscopic wavefunctions $\Psi_{l,r}$ for the left
(l) and right (r) wells thus obey ($\hbar=1$):
\begin{align}
i\frac{d\Psi_l}{dt}&=\frac{\epsilon}{2}\Psi_l-J\Psi_r+U_l|\Psi_l|^2\Psi_l+i[g-\Gamma|\Psi_l|^2]\Psi_l, \label{eq:leftwell}\\ i\frac{d\Psi_r}{dt}&=-\frac{\epsilon}{2}\Psi_r-J\Psi_l+U_r|\Psi_r|^2\Psi_r-i\gamma\Psi_r.\label{eq:rightwell}\end{align}
Here $\epsilon$ is the energy difference between the wells, including
the mean-field interaction with the exciton
reservoir.\cite{galbiati12} $J$ is the tunneling strength, and $U_{l}$
and $U_{r}$ are the matrix elements of the polariton-polariton
interaction\cite{tassone99} in the localized basis. The final two
terms in Eq. \eqref{eq:leftwell} are the net linear gain, with
coefficient $g$, and the lowest order nonlinear gain, with coefficient
$\Gamma$. For the isolated well condensation occurs for positive
$g$. In this case the linear gain term generates an exponential growth
that is stabilized by the nonlinear term. The balance of these terms
establishes a steady-state condensate with density $n_0=g/\Gamma$. We
assume that the right well is either unpumped or pumped below
threshold so that the amplitude there decays, with rate $\gamma$.

It is convenient to introduce dimensionless fields by replacing
$\Psi_{r(l)}\to \Psi_{r(l)}\sqrt{n_0}$, so that $n_0$ becomes the unit
of density. We take the corresponding blueshift $U_l n_0$ to be the
unit of energy, and factor out an overall oscillation and phase
difference between the modes by setting $\Psi_{r(l)}=e^{-i\omega t}
\psi_{r(l)} e^{(-)i\theta/2}$. Thus we obtain a dimensionless form for
the equations of motion \begin{align} i\dot{\psi_l}&=E_l\psi_l-J\psi_r
  e^{i\theta}+ig [1-|\psi_l|^2]\psi_l, \label{eq:leftwelldiml}
  \\ i\dot{\psi_r}&=E_r\psi_r-J\psi_l
  e^{-i\theta}-i\gamma\psi_r, \label{eq:rightwelldiml} \end{align}
where the energies include the mean-field shifts due to the
interactions within the condensate, \begin{align}
  E_l&=\frac{\hat{\epsilon}}{2}-\hat{\omega}+n_l,\nonumber
  \\ E_r&=-\frac{\hat{\epsilon}}{2}-\hat{\omega}+\frac{U_r}{U_l}n_r,
\label{eq:energydefs}\end{align}
and the populations are defined as $n_{l(r)}=|\psi_{l(r)}|^2$. Here
$\hat{J},\hat{\epsilon},\hat{\gamma},\hat{\omega},\hat{g}$ are
measured in units of $U_ln_0$. The dimensionless $\hat{g}$ is
$g/U_ln_0=\Gamma/U_l$, and is called $\alpha$ in
Ref.\ \onlinecite{janot13}. We note that although this parameter is a
dimensionless pump rate it is independent of the original dimensionful
pump rate in the model. It is a non-equilibrium control parameter,
which is zero in equilibrium, and large when the dissipative
nonlinearity dominates over the interactions, $\Gamma \gg U$, as in a
conventional laser.

Another theory of the driven-dissipative condensate is used in
Refs.\ \onlinecite{wouters07,wouters09,read09,read10}. In that theory
the growth of the condensate, which is due to in-scattering from a
reservoir of high-energy excitons created by the pump, is modeled by a
linear gain term. Although the nonlinear gain is not explicit it is
nonetheless present, because the linear gain coefficient depends on
the occupation of the exciton reservoir.  Solving for this reservoir
occupation on timescales for which it is time independent gives a gain
coefficient, $g(n)$, which is a function of the condensate
occupation. Expanding this as a power series and retaining the first
two terms, $g(n)=g-\Gamma n \ldots$, gives the form in
Eq. (\ref{eq:leftwell}). Note that above
threshold one must include at least the first two terms otherwise the
solution is unphysical; below threshold the second can safely be
neglected (as in Ref.\ \onlinecite{poddubny13}).

As well as the interaction within the condensate, $U$, there are also
interactions between the condensate and the population of high-energy
excitons created by the pump\cite{galbiati12}, which give rise to
energy shifts that are included in the definition of $\epsilon$. Thus
in order to explore the near-resonant case $\epsilon=0$ discussed
below one should consider a system where the bare modes are
off-resonance, so that they are brought into resonance under
asymmetrical pumping. We note that in the off-resonance case
relaxation processes provide the dominant coupling between the
wells. These are not present in our theory; we consider the case,
near-resonance, where the Josephson coupling dominates.

We have also neglected, in writing Eqs. (\ref{eq:leftwell})
and (\ref{eq:rightwell}), the spin of the polariton.\cite{shelykh10}
The spin projection of the polariton takes two values, corresponding
to right and left circular polarization states. However, condensates
typically show a high degree of polarization, generally linear
polarization in a direction determined by the crystal. In such cases
only a single polarization state is relevant, and the model can be
used to describe the spatial structure of the condensate, i.e., the
extrinsic Josephson effects. The model could be generalized to include
both polarizations, making the polariton wavefunction in each well a
spinor. In this case one can have, as well as the extrinsic effects,
intrinsic Josephson effects connected to the tunneling between the two
polarization states. These have been studied in a model without gain
and loss\cite{shelykh08}, and for the case of symmetrical
pumping\cite{read10,borgh10}.

\section{Steady-State Analysis}\label{sec3}

We now derive the phase diagram of the double-well system, examining
the steady-state solutions of Eqs. (\ref{eq:leftwelldiml}) and
(\ref{eq:rightwelldiml}), and their stability. For simplicity we
suppose that the detuning, $\epsilon$, is zero, and the interaction
strengths are identical, $U_l=U_r=U$. We consider steady-states with a
single characteristic frequency $\omega$, which corresponds to the
chemical potential in the equilibrium case. This is justified by our
numerical results, in which the steady-states are always of this
class; we have not observed steady-states with a.c. Josephson
oscillations. Setting the time derivatives in
Eqs. (\ref{eq:leftwelldiml}) and (\ref{eq:rightwelldiml}) to zero,
multiplying by $\psi_l$ and $\psi_r$, respectively, and taking real
and imaginary parts, gives: \begin{align} \hat{g}(1-n_l)n_l &
  =\hat{J}\sqrt{n_l n_r}\sin(\theta) \label{eq:ss1}\\ \hat{\gamma} n_r
  & = \hat{J}\sqrt{n_l
    n_r}\sin(\theta) \label{eq:ss2}\\ E_ln_l=(n_l-\hat{\omega})
  n_l&=\hat{J}\sqrt{n_l
    n_r}\cos(\theta) \label{eq:ss3}\\ E_rn_r=(n_r-\hat{\omega})
  n_r&=\hat{J}\sqrt{n_l n_r}\cos(\theta)\label{eq:ss4}.
\end{align} 

Eqs. (\ref{eq:ss1}) and (\ref{eq:ss2}) describe the current flows in the double-well. The term on the left of Eq. (\ref{eq:ss1}) describes the flow of polaritons from the reservoir to the left well, which is non-zero if the density there deviates from the value $n_l=1$ at which the linear gain is reduced to zero by the gain depletion. In the steady-state this current must flow as a Josephson current into the right well, which is given by the term on the right-hand side of Eq. (\ref{eq:ss1}). This Josephson current in turn accounts for the loss from the right well, as described by Eq. (\ref{eq:ss2}). Eqs. (\ref{eq:ss3}) and (\ref{eq:ss4}) are the associated pressure balance (quantum Bernoulli) equations, which state that in the steady-state the two populations must be in mechanical equilibrium through the Josephson coupling. 

Noting the equality of the right-hand sides of Eqs. (\ref{eq:ss3}) and
(\ref{eq:ss4}) we have \begin{equation}
  (n_l-\omega)n_l=(n_r-\omega)n_r. \label{eq:ssdensoromega}\end{equation}
This has a solution corresponding to the normal state, $n_l=n_r=0$. It
has two further solutions, one with $n_l=n_r$, and one with $n_l\neq
n_r$. Thus there are two classes of condensed steady-state: one in
which the density is symmetrical, as expected from the linear
eigenstates of the symmetrical double-well, and one in which it is
asymmetrical, due to the asymmetry of the pumping.

\subsection{Delocalized solutions}

We consider first the solution in which the condensate density is
equal in the two wells, $n_l=n_r$. In this case we obtain, from
Eqs. (\ref{eq:ss1}) and (\ref{eq:ss2}), \begin{equation}
  n_l=n_r=1-\frac{\gamma}{g}.\label{eq:symmleftweldens}\end{equation}
This gives a phase boundary $\gamma=g$ shown in
Fig.\ \ref{fig:phasediag1a}, which separates the normal state
(labeled N), from the symmetrical-density condensed states (labeled
$\mathrm{S}0$ and $\mathrm{S}\pi$). This phase boundary is simply the
requirement that the pumping must be sufficient to overcome the losses in
the unpumped well. A second phase boundary follows on noting that,
from Eq. (\ref{eq:ss2}), the phase difference between the wells is
given by \begin{equation}
  \gamma=J\sin(\theta).\label{eq:symmphaserel}\end{equation} Thus the
solution requires $\gamma<J$. With increasing $\gamma$ there is a
phase transition, at which this type of condensate breaks down because
the interwell current imposed by the gain and loss exceeds the
Josephson critical current.

As noted in the introduction, previous work on the incoherently-pumped
junction has focused on the case of symmetrical
pumping\cite{wouters08,eastham08,read10,borgh10}. As we find here, in
some parameter regimes there are d.c. Josepshon-like states, in which
the tunneling leads to a single condensate, and there is a fixed phase
relationship between the wells. Polarization splittings give rise,
similarly, to an intrinsic Josephson effect, fixing the polarization
direction of each mode and leading to polarization locking in extended
systems\cite{read10,borgh10}. In general such d.c. Josephson states
break down if the interwell (for the extrinsic Josephson effect) or
interpolarization (for the intrinsic effect) currents exceed the
Josephson critical current. For symmetrical pumping, this leads to a
transition to an a.c. Josephson-like
state\cite{wouters08,eastham08,borgh10}. Here we find instead that it
drives a localization transition, forming a d.c. Josephson-like state
with an imbalanced density.

In general Eq. (\ref{eq:symmphaserel}) has two solutions, so that there can be two symmetrical density condensates, with interwell phase differences in $(0,\pi/2)$ or $(\pi/2,\pi)$ respectively.  The condensate energy $\omega$ for these two solutions follows from Eq. (\ref{eq:ss3}), \begin{equation} \omega=n\mp \sqrt{J^2-\gamma^2}.\label{eq:symmdens-freq}\end{equation} For small $\gamma/J$ one solution has $\theta\approx 0$, and one has $\theta\approx \pi$.  These values correspond to those expected from the linear regime\cite{eastham08}, i.e., condensation in the symmetrical or antisymmetrical orbital. The condensate energy, Eq. (\ref{eq:symmdens-freq}), is then the expected single-particle energy shifted by the mean-field interaction. Increasing $\gamma$ introduces interwell currents, shifting the phase difference away from the limiting values of $0$ and $\pi$, and increasing (decreasing) the energy of the ground-state (excited-state) condensate.

In conventional equilibrium Josephson junctions, where $J>0$, the
$\pi$ state is unstable, since it corresponds to an energy
maximum. However, in the driven-dissipative condensate both phase states can
be stable. The stability can be determined by linearizing
Eqs. (\ref{eq:leftwelldiml}) and (\ref{eq:rightwelldiml}) about the
steady-state solution. The details are given in the Appendix, and the
results have been incorporated on the phase diagrams shown in
Figs.\ \ref{fig:phasediag1a} and \ref{fig:phasediag1b}. We see that
both phase states are stable near the onset of condensation at
$\gamma=g$, but as $g$ increases the higher-energy solution,
labeled $\mathrm{S}\pi$, eventually becomes unstable. For small $J$,
corresponding to a condensate in which the interactions dominate over
the tunneling, the $\mathrm{S}\pi$ state is restricted to a small
region near the onset of condensation (see
Fig.\ \ref{fig:phasediag1b}), whereas for larger $J$ there is a large
region where both solutions are stable (see
Fig.\ \ref{fig:phasediag1a}). Thus we predict that polariton
condensates can realize two phase states in a single junction, so long
as the interactions are not too strong compared with the tunneling.

\subsection{Asymmetrical-density solutions}

The second class of solution has $n_l\neq n_r$, and corresponds to the situation in which the condensate density occupies predominantly one of the two wells. For such a localized or trapped solution  Eq. (\ref{eq:ssdensoromega}) gives the condensate frequency  \begin{equation}\omega=n_l+n_r. \end{equation} This corresponds to the mean-field energy shift for the total occupation of the wells, and contrasts with the corresponding result for the symmetrical-density solutions, Eq. (\ref{eq:symmdens-freq}). The densities may be determined by equating the left-hand-sides of Eq. (\ref{eq:ss1}) and Eq. (\ref{eq:ss2}), giving \begin{equation} n_r=\frac{g}{\gamma}(1-n_l) n_l. \label{eq:asymmdensrel}\end{equation} Using this result, along with Eq. (\ref{eq:ss3}), in Eq. (\ref{eq:ss1}), leads to a cubic for the left-well density \begin{equation} n_l^3-n_l^2+\gamma^2 n_l  + \gamma^2 \left( \frac{J^2}{g\gamma } - 1\right)=0.\label{eq:cubicdenseq} \end{equation} The interwell phase $\theta$ is then determined by inserting the calculated densities and condensate frequency into Eqs. (\ref{eq:ss1}-\ref{eq:ss4}), so completing the solution. Note that this fully determines the interwell phase for each solution of Eq. (\ref{eq:cubicdenseq}) (up to the trivial addition of multiples of $2\pi$); in contrast to the symmetrical-density case there is only one steady-state for each $n_l \neq n_r$.

\begin{figure}
  \includegraphics{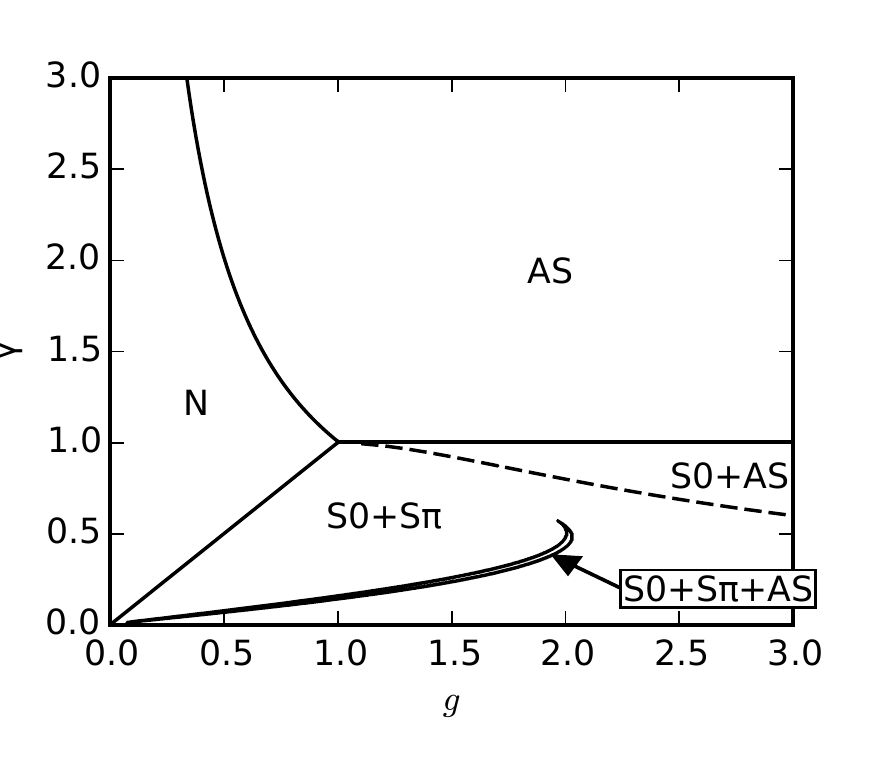}
  \caption{Phase diagram for the asymmetrically-pumped double-well
    condensate, as a function of the dimensionless pump rate $g$ for
    the left well and loss rate $\gamma$ for the right well. The
    tunneling strength $J=1$, and all energies and rates are measured
    in units of the mean-field energy shift of the left well. The
    solid and dashed lines mark phase boundaries (see text). The
    labels indicate the states present in each region: the normal
    state with no condensate (N); two condensed states with equal
    densities in the two wells, but different relative phases (S0,
    S$\pi$); and a condensed state with unequal densities in the two
    wells (AS).
  }
    \label{fig:phasediag1a}
\end{figure}

\begin{figure}
  \includegraphics{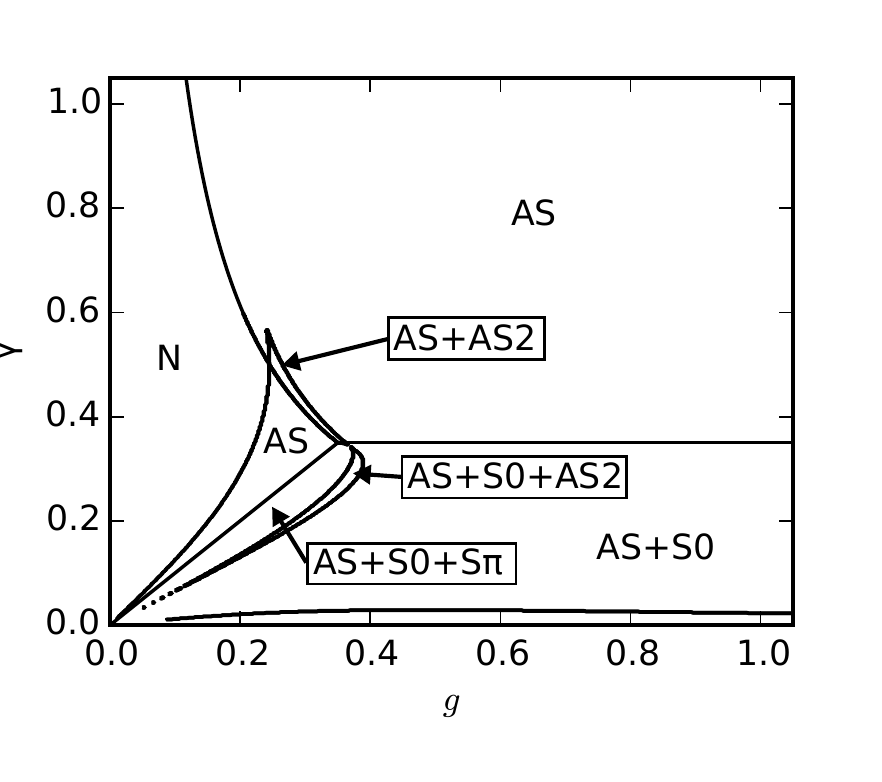}
  \caption{Phase diagram for the asymmetrically-pumped double-well condensate with tunneling strength $J=0.35$, for comparison with Fig.\ \ref{fig:phasediag1a}. The labels indicate the states
    present in each region: the normal state with no condensate (N);
    two condensed states with equal densities in the two wells, but
    different relative phases (S0, S$\pi$); and two condensed states with
    unequal densities in the two wells (AS, AS2).}
  \label{fig:phasediag1b}
\end{figure}

The cubic Eq. (\ref{eq:cubicdenseq}) implies that the asymmetrical condensate can appear continuously or at a first-order transition. The potential continuous transitions correspond to a real root becoming physical when the occupation $n_l$ crosses zero from below, which we see from the final two terms in Eq. (\ref{eq:cubicdenseq}) occurs at \begin{equation} \gamma=J^2/g.\label{eq:asymmphaseb}\end{equation} This result is plotted on the phase diagram in Fig.\ \ref{fig:phasediag1a}: it gives the continuous transition between the normal state and an asymmetrical-density condensate in the top-left quadrant. To the right of this curve there is a real-valued and positive solution for the condensate density. The potential first-order transitions occur when a pair of complex roots become real. The discriminant of the cubic, Eq. (\ref{eq:cubicdenseq}), is a quadratic in the pump parameter $g$, giving a pair of potential first-order phase boundaries, which meet at a critical point which is at $\gamma=1/\sqrt{3}$. We will discuss these results further in the next subsection.

\subsection{Phase diagram}

To complete the derivation of the phase diagram we combine the results
of the previous subsections with a linear stability analysis of the
steady-states, whose details are given in the Appendix. The results
are shown in Figs.\ \ref{fig:phasediag1a} and \ref{fig:phasediag1b},
in which the curves are the phase boundaries at which stable, physical
steady-state solutions appear or disappear.

We discuss first Fig.\ \ref{fig:phasediag1a}, which corresponds to
$J=1$, so that the tunneling strength is equal to the interaction
energy scale of the left well. We see that for $\gamma<J=1$ there is a
continuous transition, crossed as the parameter $g$ increases
from zero, from the normal state to a symmetrical-density
condensate. The phase boundary is $\gamma=g$, as predicted by
Eq. (\ref{eq:symmleftweldens}). For larger $\gamma>J=1$ the transition
is instead from the normal state to an asymmetrical-density
condensate, and follows the boundary given by
Eq. (\ref{eq:asymmphaseb}). The continuation of this curve into the
region $\gamma<J$ does not correspond to a phase boundary: The
solution which crosses zero density with increasing $g$ or $\gamma$ in
this region, and so becomes physical, does not become stable. It
becomes stable along the phase boundary shown as the dashed curve,
where also the higher energy of the two symmetrical-density solutions,
labeled $\mathrm{S}\pi$, becomes unstable. Above this is another,
horizontal, phase boundary at $\gamma=J$, as predicted by
Eq. (\ref{eq:symmphaserel}), at which the remaining
symmetrical-density solution disappears.

As noted above, further asymmetrical-density solutions are possible
when the discriminant of Eq. (\ref{eq:cubicdenseq}) is positive, so
that there are three real roots for the density. Considering also the
stability of these solutions we find, for this value of $J$, the small
curved region shown in the figure, lying inside the
symmetrical-density regime. In this curved region there is an
additional asymmetrical density solution, which appears and then
disappears with increasing $g$ at fixed $\gamma$ at the two
first-order phase boundaries shown. The phase boundary for smaller $g$
corresponds to the discriminant of Eq. (\ref{eq:cubicdenseq}) reaching
zero, and that for larger $g$ corresponds to the state becoming
unstable.

For comparison, we show in Fig.\ \ref{fig:phasediag1b} the phase
diagram for $J=0.35$, corresponding to an interaction strength
approximately three times the tunneling strength. The ranges of the
axes have been adjusted so that the ground-state symmetric-density
condensate, $\mathrm{S}0$, occupies the same area in the two
figures. As noted above, the other symmetric-density state is
suppressed by the strong interactions, and now occurs only over a
small regime just to the right of the boundary $\gamma=g$. The strong
interactions instead favor the asymmetric states: There is now a
significant area of the phase diagram where an asymmetric state
coexists with the $\mathrm{S}0$ state, and there are small regions
where two asymmetric solutions coexist. The first-order transition at
which an asymmetric state appears with increasing $g$ now lies
partially inside the normal region, and this state survives over a
much wider range of parameters.

\section{Numerical Results}\label{sec4}

\begin{figure}
\includegraphics[width=220pt]{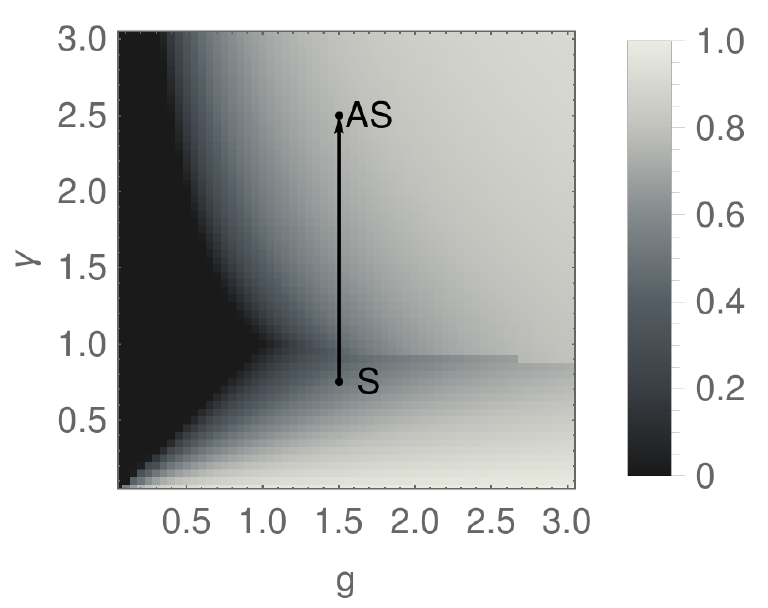}
\caption{Numerically calculated left well population for $J=1$ showing the three distinct states N, S and AS predicted in Fig.~\ref{fig:phasediag1a}. The time evolution is shown in Fig.~\ref{fig:phasediag4} for the highlighted points at $g=1.5$ for the modes S and AS corresponding to $\gamma=0.75, 2.5$ respectively.\label{fig:phasediag2}}
\end{figure}

\begin{figure}
\includegraphics[width=220pt]{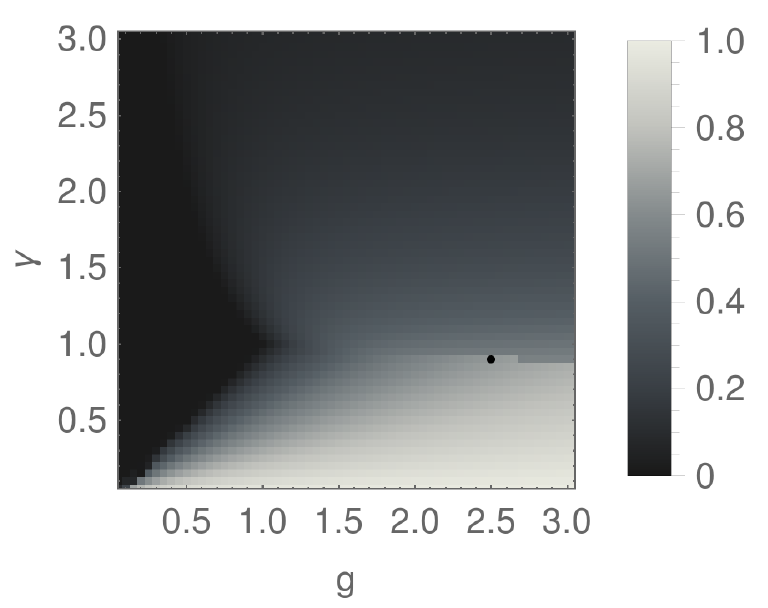}
\caption{Numerically calculated right well population for $J=1$ showing the three distinct states N, S and AS predicted in Fig.~\ref{fig:phasediag1a}. The time-dependent solution at $\gamma=0.9$ and $g=2.5$ is shown in Fig.~\ref{fig:phasediag5} in order to demonstrate control over condensate mode as a function of initial conditions.\label{fig:phasediag3}}.
\end{figure}

In this section we numerically solve Eqs. \eqref{eq:leftwell} and
\eqref{eq:rightwell} so as to study the steady-states and transient
behavior of the on-resonance double-well system. We determine the
steady-state densities by evolving from an initial state chosen to be
$\psi_l(0)=1,\psi_r(0)=0$. We take the dimensionless $J=1$ for
comparison with Fig.\ \ref{fig:phasediag1a}. The three distinct
regimes, corresponding to the normal state (N), condensation with
equal densities in the wells (S), and condensation with unequal
densities (AS), are apparent in Figs.~\ref{fig:phasediag2} and
\ref{fig:phasediag3}, which show the populations of the left and right
wells after the transient. These results are clearly consistent with
the phase boundaries obtained above. In particular we see the
transition between the normal state and an asymmetrical condensate at
$\gamma=J^2/g$, and that between the normal state and a symmetrical
condensate at $\gamma=g$. There is also a discontinuity corresponding
to switching between these solutions, close to the line
$\gamma=J$. Above this line, as argued previously, the symmetrical
solution breaks down, because the Josephson current is unable to
compensate for the unbalanced pumping. As shown in
Fig.\ \ref{fig:phasediag1a}, the asymmetric solution is in fact stable
below this line (i.e. for smaller $\gamma$) giving a region of
parameter space in which both steady-states are stable. In this
parameter regime the steady-state is selected by the initial
conditions for the condensate mode. This is consistent with the
numerical results, where the discontinuity lies in the region of phase
coexistence. 

The appearance of imbalanced densities for a condensate in a
double-well potential resembles the self-trapping of equilibrium
condensates\cite{smerzi97}. However, self-trapping in equilibrium is
an effect of interactions, and occurs only when the interaction
strength dominates over the tunneling $Un\gg J$. In contrast we see
that in non-equilibrium condensates asymmetrical densities can be
caused by an inhomogeneous pumping. Since the phase boundary
$\gamma=J$ is independent of the interaction strength, the transition
to asymmetrical condensation can be induced by increasing the
dissipation $\gamma$, even when the interactions $U$ are
negligible. The effect is instead driven by the dissipative
nonlinearity $\Gamma$.


\begin{figure}
\includegraphics[width=3in]{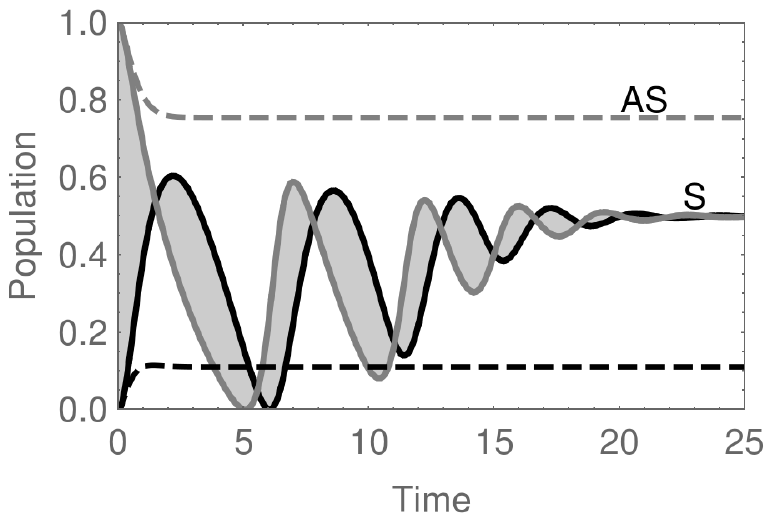}
\caption{Time dependent populations for the left (gray) and right (black) wells for the parameters corresponding to the points highlighted in Fig.~\ref{fig:phasediag2}. As loss ($\gamma$) increases the steady-state switches from a symmetric density mode (S, solid curves) to an asymmetric one (AS, dashed curves). The initial state is taken to be $n_l=1,n_r=0$.\label{fig:phasediag4}}
\end{figure}

As an example of trapping by dissipation we follow the trajectory
marked with the arrow on Fig.\ \ref{fig:phasediag2}, increasing the
dissipation through the transition at fixed $g$. In practice this
could be achieved by reducing the pumping: recall that the
dimensionless decay rate is the physical decay rate in units of the
mean-field energy shift. It might instead be achieved by using an
additional weak pumping of the right well to manipulate its decay rate
(it may also be necessary to manipulate the bare detuning to offset
the change in reservoir-induced blueshift). As the dimensionless loss
in the right well increases the steady-state transitions from a
symmetrical to an asymmetrical density profile. Thus the output of the
unpumped well can be switched by controlling its loss. The dynamics of
the populations at the two extreme values of $\gamma$ highlighted in
Fig.\ \ref{fig:phasediag2} is shown in Fig.~\ref{fig:phasediag4}. For
small $\gamma$ the population imbalance introduced in the initial
condition oscillates between the two wells, and these oscillations
decay towards a steady-state with equal densities. Close to that
steady-state these oscillations can be interpreted as due to the
Josephson plasmon of the junction, which is damped by the gain and
loss; its frequency and decay rate follows from the stability analysis
in the Appendix. With increasing $\gamma$ the oscillations disappear,
and the condensate rapidly reaches a steady-state which is trapped in
the left well. This dissipative method differs from previous
self-trapping mechanisms which use the interaction energy, $Un$, to
control the trapping of condensates\ \cite{gillet14}.

\begin{figure}[t!]
\includegraphics[width=3in]{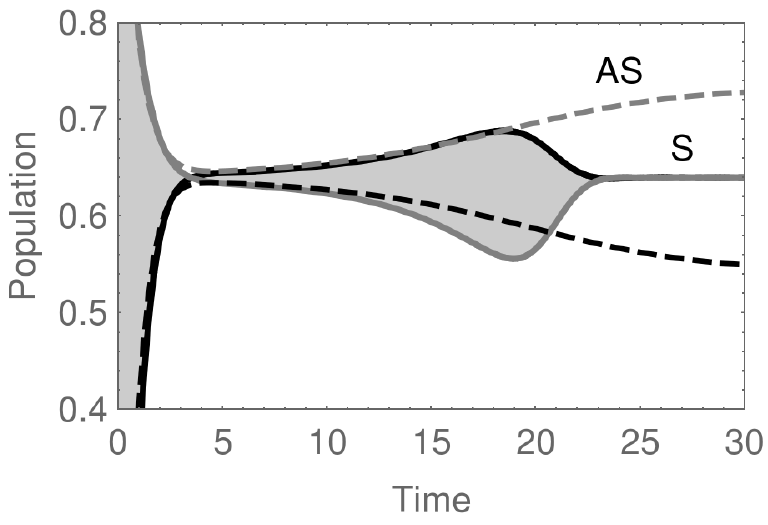}
\caption{Time dependent populations for the left (gray) and right (black) wells in the two-state region of Fig.~\ref{fig:phasediag1a}: $\gamma=0.9$ and $g=2.5$. Depending on the initial conditions either the symmetric density mode (S) or the asymmetric one (AS) is reached. The initial states have $n_l=1$ (S, solid curves) or $n_l=2$ (AS, dashed curves), and $n_r=0$.\label{fig:phasediag5}}.
\end{figure}

Self-trapping could also be induced by varying the initial conditions
for those values of $\gamma$ and $g$ where both symmetrical and
asymmetrical steady-states are stable. In these regimes (similarly to
equilibrium self-trapping\cite{raghavan99}) the initial conditions for
the condensate mode determine whether the steady-state has a
symmetrical or asymmetrical density. A specific example given in
Fig.~\ref{fig:phasediag5} demonstrates this for the highlighted point
in Fig.~\ref{fig:phasediag3}. For the corresponding values of $\gamma$
and $g$, we look at two different initial conditions for which the
left well is altered from $\psi_l(0)=1$ to a value further above
threshold $\psi_l(0)=2$. In the latter case, the Josephson current
does not become sufficient to populate the wells symmetrically, and we
see the steady-state mode change from the symmetric to asymmetric
one. One can also switch between these steady-states, as well as the
different phase states of the symmetrical condensate, by driving the
wells externally. This could be used to implement optical switches and
memories, with the desirable feature that the output of one switch
(the intensity from one of the wells) can form the input of
another. The possibility of manipulating the initial conditions so as
to induce self-trapping for polariton condensates has already been
shown experimentally in Ref.\ \onlinecite{abbarchi13}, considering
self-trapping as a transient phenomenon in the absence of gain. The
present work shows how such concepts can be extended to apply to the
steady-states of driven-dissipative condensates.

In experiments the initial conditions for a driven-dissipative
condensate might be controlled by using a resonant seed pulse to
prepare the initial state. If the pump is then rapidly switched on
then the state selected will be determined by these initial
conditions. An alternative possibility would be to slowly increase the
pumping, in which case we expect the system to adiabatically follow a
particular state in the phase diagram. Large changes in pumping,
however, would require a corresponding change in the bare detuning, in
order to maintain the modes near resonance. In practice it would not
be possible to keep the modes exactly resonant, as assumed
above. However, we have checked numerically that this is not
essential. We find that for small non-zero detunings the transition
from a symmetrical to asymmetrical density mode survives, although it
becomes one from a weakly to a strongly asymmetrical mode.


\section{Summary}

To summarize, we have studied the steady-states and dynamics of a
two-mode polariton condensate, such as the Josephson junction in a
double-well potential, for the case where one mode is pumped. We have
shown that such asymmetrical pumping provides a mechanism by which
condensates can become localized in one well which, in contrast to
conventional self-trapping, does not rely on strong interactions,
being caused instead by nonlinear gain. This will help open up the
study of condensate localization and self-trapping in
driven-dissipative condensates, which we predict can occur in both
strongly and weakly interacting regimes. As we have shown, a simple
two-mode polariton condensate supports a variety of coherent
steady-states, distinguished by both their density and phase
profiles. It would be interesting to extend our work to systems such
as lattices of junctions, and so explore the nonlinear dynamics of
spatially extended driven-dissipative condensates. Another interesting
direction would be to consider the use of time-dependent parameters to
control the trapping\cite{cui10,wang06}. The possibility of realizing
multiple steady-states in a two-mode condensate, and switching between
these states with applied fields, may be useful for optical switches
and memories.

\begin{acknowledgments}
PRE acknowledges support from Science Foundation Ireland (SFI) grant
15/IACA/3402. FB acknowledges support from SFI grant 15/IFB/3317 and
the SFI funded center CONNECT grant 12/RC/2276.
  \end{acknowledgments}

\appendix
\section{Stability analysis}

In the main text we present results for the stability of the steady-states of Eqs.\ (\ref{eq:leftwelldiml},\ref{eq:rightwelldiml}). This is determined by setting $\psi_{l(r)}(t)=\psi_{l(r)}^{0}+\delta_{l(r)}(t)$, where $\psi_{l(r)}^{0}$ are the steady-state fields. The equations-of-motion, to first order in $\delta$, can be written in the block form
\begin{equation} i \frac{d}{dt}\begin{pmatrix} \eta \\ \eta^\ast \label{eq:linearizedeqs}\end{pmatrix}
=\begin{pmatrix} A & B \\ -B^\ast & -A^\ast \end{pmatrix}\begin{pmatrix} \eta \\ \eta^\ast\end{pmatrix}=M\begin{pmatrix} \eta \\ \eta^\ast \end{pmatrix},\end{equation} where \begin{align*} \eta&=\begin{pmatrix} \delta_l \\ \delta_r\end{pmatrix}, \\ 
A&=\begin{pmatrix} 
-\omega + 2 n_l + ig(1-n_l) & -Je^{i\theta} \\
-Je^{-i\theta} & -\omega + 2 n_r - i\gamma \end{pmatrix}, \\
B&=\begin{pmatrix}
n_l(1-ig) & 0 \\
0 & n_r \end{pmatrix}.
\end{align*} Here $n_{l(r)}$ denotes the steady-state densities and $\theta$ the steady-state phase difference, obtained by solving Eqs. (\ref{eq:ss1}-\ref{eq:ss4}), and, as in those equations, we have set $\epsilon=0, U_r=U_l$. We see that the solutions of Eq. (\ref{eq:linearizedeqs}) are of the form $\eta=\eta_0 e^{-i\nu t}$, where the complex frequencies $\nu$ are the eigenvalues of the matrix $M$. One eigenvalue has zero decay rate and frequency, and corresponds to the expected (undamped) phase mode of the condensate. The remaining eigenvalues describe the excitation spectrum in terms of the frequencies $\Re(\nu)$, and decay rates, $\lambda=-\Im(\nu)$ of small fluctuations; the latter are positive for a stable steady-state.

\bibliographystyle{apsrev}

\end{document}